\documentclass[%
preprint,
superscriptaddress,
notitlepage,
 amsmath,amssymb,
prb,
floatfix,
]{revtex4-2}

\usepackage{etoolbox}

\makeatletter
\patchcmd{\frontmatter@abstract@produce}
  {\vskip200\p@\@plus1fil
   \penalty-200\relax
   \vskip-200\p@\@plus-1fil}
  {}
  {}
  {}
\makeatother

\usepackage{graphicx}
\graphicspath{{../Figures/}}
\usepackage{dcolumn}
\usepackage{bm}
\usepackage{appendix}
\usepackage[
    colorlinks=true,
	citecolor=blue,
	linkcolor=blue,
	urlcolor=blue]{hyperref}

\usepackage{cleveref}

\usepackage{subcaption}
\usepackage{xcolor}

\newcommand{\dfdx}[2]{\frac{\mathrm{d}#1}{\mathrm{d}#2}}

\newcommand{\ket}[1]{|#1\rangle}
\newcommand{\bra}[1]{\langle#1|}
\newcommand{\braket}[2]{\langle#1|#2\rangle}

\newcommand{\cc}[1]{\overline{#1}}
\newcommand{\transpose}[1]{#1^\top}

\newcommand{\bigo}[1]{\mathcal{O}\left(#1\right)}

\newcommand{\qexpect}[1]{\langle#1\rangle}

\newcommand{\tr}[1]{\mathrm{Tr}\left(#1\right)}

\newcommand{\dnorm}[1]{\left\lVert#1\right\rVert}

\newcommand{\real}[1]{\mathrm{Re}\left[#1\right]}
\newcommand{\imag}[1]{\mathrm{Im}\left[#1\right]}

\newcommand{\identity}{\mathbb{I}}

\begin{document}


\title{Digital quantum simulation of open quantum systems using quantum imaginary time evolution}

\author{Hirsh Kamakari}
\affiliation{Division of Engineering and Applied Science, California Institute of Technology, Pasadena, CA 91125, USA}

\author{Shi-Ning Sun}
\affiliation{Division of Engineering and Applied Science, California Institute of Technology, Pasadena, CA 91125, USA}

\author{Mario Motta}
\affiliation{IBM Quantum, IBM Research Almaden, San Jose, CA 95120, USA}

\author{Austin J. Minnich}
\email{aminnich@caltech.edu}
\affiliation{Division of Engineering and Applied Science, California Institute of Technology, Pasadena, CA 91125, USA}

\date{\today}

\begin{abstract}
Quantum simulation on emerging quantum hardware is a topic of intense interest. While many studies focus on computing ground state properties or simulating unitary dynamics of closed systems, open quantum systems are an interesting target of study owing to their ubiquity and rich physical behavior. However, their non-unitary dynamics are also not natural to simulate on digital quantum devices.  Here, we report  algorithms for the digital quantum simulation of the dynamics of open quantum systems governed by a Lindblad equation using adaptations of the  quantum imaginary time evolution (QITE) algorithm.  We demonstrate the algorithms on IBM Quantum's hardware with simulations of the spontaneous emission of a two level system and the dissipative transverse field Ising model. Our work advances efforts to simulate the dynamics of open quantum systems on quantum hardware.
\end{abstract}

\maketitle

\clearpage
\section{\label{sec:intro}Introduction}







The development of quantum algorithms to simulate the dynamics of quantum many-body systems is now a topic of interest owing to advances in quantum hardware \cite{cerezo2020variational, georgescu2014quantum, bharti2021noisy}. While the real-time evolution of closed quantum systems on digital quantum computers has been extensively studied in the context of spin models \cite{fauseweh2020digital, smith2019simulating, chiesa2019quantum, lamm2018simulation, endo2020, cirstoiu2020variational, gibbs2021long}, fermionic systems \cite{barends2015digital, arute2020observation}, electron-phonon interactions \cite{macridin2018a}, and quantum field theories  \cite{jordan2012quantum, kharzeev2020real, kreshchuk2020quantum}, fewer studies have considered the time evolution of open quantum systems, which  exhibit rich dynamical behavior  due to coupling of  the system to its environment \cite{lidar2019lecture, breuer2002}. However, this coupling leads to non-unitary evolution which is not naturally simulable on quantum hardware. 

Early approaches to overcome this challenge included use of the quantum simulators' intrinsic decoherence \cite{tseng2000} and direct simulation of the environment \cite{wang2011, su2020, cattaneo2021}. Theoretical works examined  the resources required for efficient quantum simulation of Markovian dynamics \cite{bacon2001, sweke2015, kliesch2011}, concluding that arbitrary quantum channels can be efficiently simulated by combining elementary quantum channels. Recently, several algorithms have been proposed for the digital quantum simulation of open quantum systems on the basis of the Kraus decomposition of quantum channels \cite{wei2016, hu2020, hubisz2020quantum, hu2021general, head2021, delredriven2020} as well as variational descriptions of general processes  to simulate the stochastic Schr\"odinger equation \cite{endo2020, cerezo2020variational} and the Lindblad equation \cite{haug2020generalized}. Recently, explicit Trotterization of the Lindblad equation was used to simulate damping and dephasing of a single qubit using an additional ancilla qubit \cite{han2021experimental}.  

Simulation via Kraus decomposition is convenient when the Kraus operators corresponding to the time evolution of the system are known, such as modelling decoherence with amplitude damping or depolarizing channels. However, determining the Kraus operators of a general system  requires either computing the full unitary evolution of both the system and environment or casting a master equation into an operator sum representation for the density operator. The latter procedure can be approximated analogously to Trotterization \cite{delredriven2020, hu2021general} but requires either reset of ancillae qubits or a qubit overhead which scales linearly with the number of time steps in the simulation. Exactly determining the Kraus operators from the Lindblad equation is a classically hard task which is equivalent to solving the master equation \cite{andersson2007} and so can only be applied to small systems.  Explicit Trotterization circumvents the need to determine the Kraus operators representing the time evolution but has the same ancilla qubit overhead as in as the Kraus decomposition methods.   Variational approaches \cite{endo2020, cerezo2020variational, yoshioka2020} offer an alternative for simulating open system dynamics, but as in the case of closed systems require an ansatz and a potentially high dimensional classical optimization which is an NP-hard problem \cite{bittel2021training}.   A quantum simulation of the stochastic Schr\"odinger equation was emulated in Ref.~\cite{endo2020}. In this case, the quantum jumps, or discontinuous changes in the quantum state, was implemented via variational matrix-vector multiplication, thus incurring the  disadvantages previously mentioned for  variational approaches.   

The common feature of the above algorithms is that they reformulate non-unitary open system dynamics into unitary dynamics which can be simulated on a quantum computer. A similar approach is used in variational approaches to imaginary time evolution \cite{mcArdle2019} and the quantum imaginary time evolution (QITE) algorithm, which has recently been introduced as a way to prepare ground states and compute thermal averages on near-term devices \cite{motta2020}. QITE has since been used to compute finite-temperature correlation functions of many-body systems \cite{sun2021}, scattering in the Ising model \cite{yeter2020a}, and binding energies in quantum chemistry \cite{gomes2020, yeter2020b} and nuclear physics \cite{yeter2020b}. It is therefore natural to consider how QITE might be adapted for open quantum system evolution.

Here, we report quantum algorithms to simulate open quantum dynamics using adaptations of the QITE algorithm and demonstrate them on IBM Quantum hardware. The first algorithm casts the Lindblad equation for the density operator into a Schr\"odinger-type equation with a non-Hermitian Hamiltonian. Time evolution is then achieved by simulating the unitary evolution via Trotterization, corresponding to the Hermitian component of the Hamiltonian and using QITE to simulate the anti-Hermitian component of the Hamiltonian. The second algorithm expresses the density operator in terms of an ansatz which is preserved during both real and imaginary time evolution. We demonstrate these algorithms on IBM Quantum hardware for two cases: the spontaneous emission of a two level system (TLS) in a heat bath at zero temperature, and the dissipative transverse field Ising model (TFIM) on two sites. We observe good agreement between the exact and hardware results, showing that the dynamics of open quantum systems are accessible on near-term quantum hardware.

\section{Theory}
The dynamics of a Markovian open quantum system can be described by the Lindblad equation

\begin{equation}
\dfdx{\rho}{t} = - i [ H, \rho] + \sum_k\left( L_k \rho L_k^\dagger - \frac{1}{2}\{ L_k^\dagger L_k , \rho \}\right)
\end{equation}
where $\rho$ is the density operator of the system, $H$ is the system's Hamiltonian, and $L_k$ are operators describing the coupling to the environment. 
The master equation in Lindblad form is often derived assuming weak coupling between system and environment and absence of memory effects (Born-Markov approximation) \cite{lindblad1976generators,breuer2002}.

We present two algorithms to simulate the master equation in Lindblad form on a digital quantum computer. The first quantum algorithm, based on a vectorization of the density operator, is described in Sec.~\ref{sec:algo_a};  the second algorithm, which combines a QITE adaptation with an ansatz for the time-dependent density operator, is presented in Sec.~\ref{sec:algo_b}.

\begin{figure}
  \centering
  {\includegraphics[width=\linewidth]{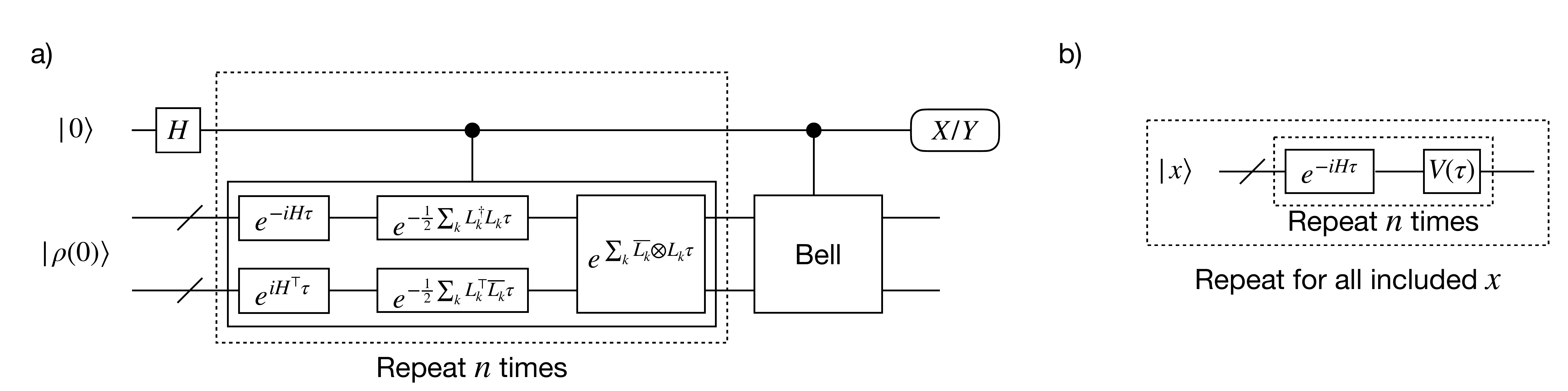}
  \phantomsubcaption\label{fig:1a}
  \phantomsubcaption\label{fig:1b}}
  \caption{Circuit diagrams for Trotterized time evolution of the density operator with $N$ Trotter steps. a) Time evolution for the vectorized density operator $|\rho\rangle$ (Algorithm I). $e^{-iH_1\tau}$ is a unitary operator and can be directly implemented on the quantum simulator. The non-unitary terms $e^{-\frac{1}{2}\sum_k L_k^\dagger L_k\tau}$ and $e^{\sum_k \cc{L_k}\otimes L_k \tau}$ are implemented via QITE. The unitary labelled ``Bell'' represents a unitary preparing the generalized $2n$-qubit Bell state. b) Time evolution for the purification-based algorithm (Algorithm II). $x$ is a bit-string included in the index set $I$. $V(\tau)$ represents the non-unitary terms which need to be applied to the system for a time step $\tau$.  In both figures / denotes a bundle of $n$ qubits.}
  \label{fig:circuit_diagrams}
\end{figure}

\subsection{Algorithm I}
\label{sec:algo_a}

The Lindblad equation can be rewritten as a Schr\"odinger-type equation with a non-Hermitian Hamiltonian by transforming the $2^n\times 2^n$ density operator $\rho$ into an $4^n$ component vector $\ket{\rho}$ by column stacking the density operator \cite{havel2003}. The resulting transformation of the Lindblad equation is 

\begin{equation}
\dfdx{\ket{\rho}}{t} = \left[-i\identity\otimes H + iH^\top\otimes\identity + \sum_{k}(\cc{L_k}\otimes L_k -\frac{1}{2}\identity\otimes(L_k^\dagger L_k) -\frac{1}{2}(L_k^\top \cc{L_k})\otimes\identity)\right]\ket{\rho}
\label{eq:vectorized_lme}
\end{equation} 
where the bar indicates entry-wise complex conjugation and $\ket{\rho}=\ket{\rho(t)}$ is the vectorized density operator \cite{havel2003}. Separating Eq.~\ref{eq:vectorized_lme} into Hermitian and anti-Hermitian parts, the time evolution of the initial state can be written as:

\begin{equation}
\ket{\rho(t)} = \exp{(-i(H_1-iH_2)t)}\ket{\rho(0)} = \left[\exp{(-iH_1\tau)}\exp{(-H_2\tau)}\right]^N\ket{\rho(0)} + \bigo{\tau^2N}.
\label{eq:trotterized}
\end{equation}

\noindent
where in the last equality we have Trotterized to first order with time step $\tau = t/N$, and $H_1$ and $iH_2$ are the Hermitian and anit-Hermitian components of the vectorized Hamiltonian, respectively. The first term $\exp{(-iH_1\tau)}$ is unitary and can implemented on a quantum simulator via Trotterization and standard quantum simulation techniques \cite{nielsen2002quantum, georgescu2014quantum, lloyd1996}. The term $\exp{(-H_2\tau)}$ is non-unitary and so cannot be directly applied to the quantum register. Instead, we implement it on a digital quantum simulator via analogy to quantum algorithms for imaginary time evolution \cite{motta2020}.

Imaginary time evolution of the Schr\"odinger equation with  Hamiltonian $H$ is carried out formally by substituting $\beta = it$ into the real time propagator $\exp{(-itH)}$. This technique is typically used to find ground states $\ket{\psi}=\lim_{\beta\rightarrow\infty} \ket{\phi(\beta)}/||\ket{\phi(\beta)}||$, where $\ket{\phi(\beta)}=\exp{(-\beta H)}\ket{\phi(0)}$ and $\ket{\phi(0)}$ has non-zero overlap with a ground state. If we interpret $H_2$ as the Hamiltonian of a system in the extended Hilbert space, $\exp{(-H_2\tau)}$ is an imaginary time evolution operator generated by $H_2$. The full time evolution is then applied as sequence of real and imaginary time evolutions, as shown in Fig.~\ref{fig:1a}.

We present a brief review of the QITE algorithm reported in Ref.~\cite{motta2020} which is used as a subroutine in this work. The QITE algorithm  represents normalized imaginary time evolution in terms of unitary evolution as:

\begin{equation}
    \frac{e^{-\beta H}\ket{\psi}}{||e^{-\beta H}\ket{\psi}||}=e^{-i A}\ket{\psi},
\end{equation}

\noindent where $H$ is the system Hamiltonian, $\beta$ is the imaginary time, and $A$ is a Hermitian operator. The operator $A$ can be represented with real coefficients in a complete basis of Hermitian operators, typically chosen to be the Pauli strings $\sigma_i$ over the qubits of the system:

\begin{equation}
A=\sum_ia_i\sigma_i.
\end{equation}

For an imaginary time step $\beta$, the coefficients $a_i$ are determined (up to order $\beta^2$) by the linear system $Sa=b$, with 
\begin{equation}
\begin{cases}
    S_{ij}=\langle\psi|\sigma_i^\dagger \sigma_j|\psi\rangle,\\
    b_i = \frac{-i}{\sqrt{c}}\langle\psi|\sigma_i^\dagger H|\psi\rangle
\end{cases}
\end{equation}
where $c=\qexpect{\exp{(-2\beta H)}}$ is the norm squared of the un-normalized imaginary time evolved state.

Once the desired time and state $\ket{\rho(t)}$ are reached, measurements of an observable $O$ are obtained by evaluating the expectation value $\qexpect{O}(t) = \tr{O\rho(t)}$ as $\braket{O^\dagger}{\rho}$. $\ket{O}$ is the vector obtained from column stacking the matrix representation of $O$   and so only the matrix representation of $O$ in the computational basis is needed for this step . Lindbladian evolution preserves $\tr{\rho}$ whereas the algorithm preserves $\tr{\rho^2}=\braket{\rho}{\rho}$,   meaning that   the operator $\rho$ obtained from matricizing $\ket{\rho(t)}$ is not strictly a density matrix. However, the final state can be renormalized to have unit trace as $\rho'(t)=\rho(t)/\tr{\rho(t)}$. In practice, we normalize the final expectation value of a given observable instead.   The final physical observables are thus given by $\qexpect{O}/\tr{\rho}$. Therefore, obtaining measurements of observables on the state requires evaluating  both $\qexpect{O}$ and $\tr{\rho}$ at each time step.

Both quantities $\qexpect{O}(t)$ and $\tr{\rho(t)}$ can be obtained using a Hadamard test circuit \cite{somma2002simulating}. In particular, $\tr{\rho}$ can be evaluated up to a prefactor of $2^{-n/2}P^{-1/2}$ as $\bra{0} V^\dagger U \ket{0}$, where $U$ is the circuit that prepares $\ket{\rho}$, $V$ prepares the generalized Bell state $\ket{\beta}=2^{-n/2}\sum_x\ket{x}\otimes\ket{x}$, $\ket{x}$ are the computational basis states on $n$ qubits, and $P$ is the purity of the initial state. Preparing the $2n$ qubit Bell state requires $n$ Hadamard  and $n$ CNOT gates. Assuming $\ket{\rho}=U\ket{0}$ for a unitary $U$ with gate decomposition requiring $u_1$ and $u_2$ single-qubit and CNOT gates, respectively, the measurement of $\tr{\rho}$ requires a circuit with $\bigo{n+u_1}$ single-qubit gates, $\bigo{n+u_1}$ CNOT gates, and $\bigo{n+u_2}$ CCNOT gates. 

Measurement of $k-$local observables can be carried out similarly. We assume here without loss of generality that the $k-$local observable $O$ has support on the first $k$ qubits. The vectorized state can then be written as $\ket{\rho} = P^{-1/2}\sum_{x_1,x_2,y_1,y_2}\rho_{x_1x_2y_1y_2}\ket{x_1x_2y_1y_2}$ where $x_1,y_1$ and $x_2,y_2$ are length $k$ and $(n-k)$ bit strings, respectively, and $P$ is the purity of the initial state. Defining the state 

\begin{equation}
\ket{O^\dagger}=\sum_{x_1y_1z}\frac{\cc{O_{x_1y_1}}}{\sqrt{2^{n-k}\tr{O^\dagger O}}}\ket{x_1zy_1z},
\end{equation}

\noindent where the over-bar indicates complex conjugation, the expectation value of $O$ can be evaluated (up to a pre-factor) as 

\begin{equation}
    \braket{O^\dagger}{\rho}=\sum_{x_1y_1z}\frac{O_{x_1y_1}}{\sqrt{2^{n-k}}}\frac{\rho_{x_1zy_1z}}{\sqrt{P}}=\frac{\tr{O\rho}}{\sqrt{2^{n-k}\tr{O^\dagger O}P}}.
\end{equation}

The state $\ket{O^\dagger}$ can be prepared as 

\begin{equation}
    U_{O^\dagger}V_{n-k}\ket{0_k,0_{n-k},0_k,0_{n-k}}=U_{O^\dagger}\left[ \frac{1}{\sqrt{2^{n-k}}}\sum_z \ket{0_k,z,0_k,z}\right]
\end{equation}

\noindent where $V_{n-k}$ prepares the $n-k$ generalized Bell state and $U_{O^\dagger}$ prepares the $2k$ qubit state

\begin{equation}
\ket{O^\dagger}=\sum_{x_1y_1}\frac{\cc{O_{x_1y_1}}}{\sqrt{\tr{O^\dagger O}}}\ket{x_1y_1}.
\end{equation}

We then measure the un-normalized expectation value of $O$ using the Hadamard test. Since the purity is conserved by the algorithm, all observables can be renormalized after the measurement. Assuming a decomposition of $U$ into $u_1$ and $u_2$ single-qubit and CNOT gates, respectively, and $V$ into $v_1$ and $v_2$ single-qubit and CNOT gates, the total overhead for measurement of observables (including the trace evaluation) is $\bigo{n + u_1+v_1}$ single-qubit gates, $\bigo{n + u_1+v_1}$ CNOT gates, and $\bigo{n + u_2+v_2}$ CCNOT gates.

\subsection{Algorithm II}
\label{sec:algo_b}

Algorithm I allows for efficient simulation of the full density operator for many physical systems characterized by local interactions; however, it requires a doubling of the number of qubits and an overhead of an ancilla and controlled operations for evaluating observables. In particular, the circuit required for measurements is too deep for near-term hardware. We therefore introduce a second algorithm  based on the variational ansatz used to obtain the non-equilibrium steady states of Markovian systems \cite{yoshioka2020,endo2021hybrid} that overcomes these limitations. The isomorphism maps a density operator as

\begin{equation}
    \rho=\sum_{x\in I}p_xU\ket{x}\bra{x}U^\dagger\rightarrow\ket{\rho}=\sum_{x\in I}p_xU\ket{x}\otimes \cc{U}\ket{x}.
\end{equation}
where the $\ket{x}$'s label the $n$-qubit computational basis states and $I$ is a subset of all $2^n$ possible bit-strings. In the rest of the paper the index set $I$ is implied. We note that although we are using an ansatz for this algorithm, any density operator can be represented in this form provided the index set $I$ is large enough.   However, it should be noted that assuming polynomial resources to store the bit-string weights implies that the present algorithm employs a sparse approximation to represent the density matrix.  

The Lindblad master equation is mapped identically to the vectorization mapping, resulting in Eq.~(\ref{eq:vectorized_lme}). The propagator is again Trotterized and each term can be applied term by term. The unitary part of the propagator preserves the ansatz, as 

\begin{equation}
    \exp{\left(\left(-i\identity\otimes H + iH^\top\otimes\identity\right)\tau\right)}\sum_xp_xU\ket{x}\otimes \cc{U}\ket{x}=\sum_xp_xe^{i\transpose{H}}U\ket{x}\otimes e^{-iH}\cc{U}\ket{x}
\end{equation}
and $\cc{e^{-iH}}=e^{i\transpose{H}}$ for Hermitian $H$. The remaining terms in the Trotterized propagator are of the form $\exp{(-L_k^\top\cc{L_k}\tau/2)}\otimes\exp{(-L_k^\dagger L_k\tau/2)}$ and $\exp{(\cc{L_k}\otimes L_k\tau)}$. The first term preserves the ansatz but is non-unitary, while the second term does not preserve the ansatz and is non-unitary. Considering the first of non-unitary terms, as in the original QITE algorithm we seek a set of numbers $q_x$ and a Hermitian operator $A$ such that

\begin{equation}
V_k\sum_xp_xU\ket{x}\otimes \cc{U}\ket{x}=\sum_x(p_x+q_x)\exp{(iA)}U\ket{x}\otimes \exp{(-i\cc{A})}\cc{U}\ket{x} + \bigo{\tau^2},
\end{equation}
where $V_k=\exp{(-\tau L_k^\top\cc{L_k}/2)}\otimes\exp{(-\tau L_k^\dagger L_k/2)}$. As shown in Section I the Supplementary Materials, we find that  

\begin{equation}
    q_x=-\tau p_x\real{\bra{x}U^\dagger \transpose{L_k}\cc{L_k}U\ket{x}}.
\end{equation}

Decomposing $A$ into a weighted sum of Pauli strings, $A=\sum_ja_j\sigma_j$, we find that the coefficients $a_j$ satisfy the linear system $Sa=b$, with

\begin{align}
    S_{ij}&=\sum_xp_x^2\real{\bra{x}U^\dagger(\sigma_i\sigma_j+\sigma_j\sigma_i)U\ket{x}}-2\sum_{xy}p_xp_y\real{\bra{x}U^\dagger\sigma_iU\ket{y}\bra{x}U^\dagger\sigma_jU\ket{y}},\\
    b_i&=-\tau\left(\sum_xp_x^2\imag{\bra{x}U^\dagger \sigma_i\transpose{L_k}\cc{L_k}U\ket{x}}+\sum_{xy}p_xp_y\imag{\bra{x}U^\dagger\sigma_iU\ket{y}\bra{y}U^\dagger \transpose{L_k}\cc{L_k} U\ket{x}}\right).
\end{align}

\noindent The elements $q_x$, $S$, and $b$ for the second non-unitary term, $\exp{(\cc{L_k}\otimes L_k\tau)}$, take a similar form. The derivation for both terms is given in Section I of the Supplementary Materials. We note that the total probability $\sum_x p_x = 1$ is conserved by the algorithm since $\sum_x q_x=0$ at each time step as shown in Eq.~(S19).

With this ansatz, observables are computed as $\qexpect{O}=\sum_xp_x\bra{x}U^\dagger OU\ket{x}$, requiring the propagation of each $\ket{x}$ in parallel while storing the $p_x$'s. The final observable is computed as a classical average over all the propagated states and $p_x$. It is important to note that although the ansatz lies in a dilated Hilbert space, all measurements take place on the original system and no entangling operations between the system and ancilla are needed, and so no ancillae qubits are needed. In particular, for each time step measurements on the original Hilbert space are used to determine the Hermitian matrix $A$.  Expectation values of observables on this state are computed using the standard methods \cite{georgescu2014quantum, cerezo2020variational}. 

The benefits of Algorithm II are that it requires no ancilla qubits, and no Hadamard test is required for measurements of observables. These characteristics, trading quantum for classical resources and simulating large quantum circuits using smaller quantum computers are important for near-term hardware \cite{bravyi2016trading, peng2020simulating,gujarati2021reducing, eddins2021reducing, eddins2021inprep}.   In particular, Algorithm II allows for halving the number of required qubits as in Ref.  \cite{eddins2021inprep}, allowing simulation of larger physical systems by increasing the classical and quantum computational time  while decreasing the required number of qubits.   Its  drawbacks are the sparse representation of the density matrix and  the number of measurements required to evolve the system.   We discuss this overhead in the following section. 

\section{\label{sec:runtime} Run time bounds, computational overheads, and errors}

In this section, we discuss the run times, quantum and classical computational overheads, and errors associated with each algorithm. Other sources of errors, such as those associated with noisy hardware, are not addressed here as they are non-algorithmic errors.

\subsection{Run time bounds}
We first bound the run time of Algorithm I. For each time step in the Trotterization, the algorithm requires applying the imaginary time propagator $\exp{(-H_2\tau)}$, where $\tau=t/N$ and $N$ is the number of Trotter steps for the time evolution. Assuming a local Lindblad equation, $H_2$ is a sum of $m_2$ local terms $h_l$ such that $H_2=\sum_{l=1}^{m_2}h_l$, where $m_2$ scales polynomially with system size. The imaginary time evolution operator $\exp{(-H_2\tau)}$ is implemented by additional Trotterization. For a given desired error $\epsilon_2$, we Trotterize the imaginary time evolution into $p_2$ steps.  From Eq.~(3.8) of Ref.~\cite{suzuki1976generalized}, we find that for $p_2>1/\epsilon_2$ the error in the $p_2$-step approximation is bounded by $\epsilon_2$, assuming  the number of Trotterization steps for time evolution $N$ is sufficiently large that $3 m_2 t v_2 / N < 1$, where $v_2=\max_l\{\dnorm{h_l}\}$. 

Each term in the Trotterization is an imaginary time increment and so corresponds to a rotation by a unitary operator supported on $D$ qubits where $D$ is the domain size. An arbitrary $D$ qubit unitary can be decomposed exactly into $\bigo{D^2 4^D}$ single-qubit and CNOT gates \cite{nielsen2002quantum}. The total contribution to running time from all the imaginary time evolutions is $\bigo{N m_2 D^2 4^D/\epsilon_2}$.

Algorithm I also has additional unitaries $\exp{(-i H_1 \tau)}$ interleaved between each QITE step, leading to an additional overhead. Because $H_1$ is a sum of local terms, $H_1= \sum_{l=1}^{m_1} h_l$, $\exp{(-i H_1 \tau)}$ needs to be  Trotterized as well. Performing a similar analysis for the real time evolution, we find the total running time to be 

\begin{equation}
    T= \bigo{N m_1 k^2 4^k/\epsilon_1 + N m_2 D^2 4^D/\epsilon_2},
    \label{eq:asymptotic_complexity}
\end{equation}
where $\epsilon_1$ is the allowable Trotter error for the real-time evolution and $k$ is the maximum number of qubits acted on by each term in the Hamiltonian. In the first term on the right hand side, we have assumed that each $k-$local unitary can be exactly decomposed into $\bigo{k^2 4^k}$ single qubit and CNOT gates \cite{nielsen2002quantum}.

A similar analysis can be carried out for Algorithm II, resulting in the same run-time up to constant factors with the following difference. The errors appearing in the run-time bound for Algorithm II do not include errors incurred from approximating the density operator with a strict subset of all bit-strings.  Although in principle any density operator can be represented by the sum $\sum_x p_x U\ket{x}\otimes \cc{U}\ket{x}$, this sum contains exponentially many terms and so only a subset of all possible bit strings can be included efficiently. Exclusion of bit-strings leads to an error in representing the state given by $\sum_{x\in I^c}p_x$, where $I$ is an index set containing all bit strings to be included, and $I^c$ is its complement. In practice, this error would have to be assessed by stochastically sampling the bit-strings until the simulation converges. 

\subsection{Measurement and classical computational overheads}

Provided that the finite domain approximation holds, the largest computational overhead (apart from running time) of both algorithms is the measurement overhead. For Algorithm I, this measurement overhead is the same as in the original QITE algorithm. State tomography over each domain consisting of D qubits needs to be carried out to construct the unitaries over that domain, requiring $\bigo{4^D}$ measurements. Assuming a 1-dimensional lattice, there are $\bigo{n/D}$ domains, and so the total measurement overhead is $\bigo{(n/D)4^D}$ per time step. Similar bounds can be obtained for lattices in higher dimensions.

Algorithm II requires measurement of the matrix elements $\bra{x}U^\dagger\sigma_iU\ket{y}$ for all Pauli strings $\sigma_i$ supported on a domain $D$ (measured in qubits) and all bit strings in $x,y\in I$ for some subset $I$ of the $2^n$ $n$-bit strings. Measuring all matrix elements necessitates running $\bigo{L4^D|I|^2}$ circuits per time step, where $L$ is the number of Lindblad operators on the domain and $|I|$ is the number of bit-strings  included in the computation. For the algorithm to be efficient, the number of bit strings included in $I$ must scale polynomially or slower with system size.

The finite-domain approximation required from QITE is accurate in many cases because the domain size $D$ can generally be taken to be smaller for dissipative systems compared to the same system with no dissipation, as dissipation generally reduces a system's correlation length \cite{kastoryano2013rapid}.  It should be noted that a reduced correlation length that decreases the cost for quantum algorithms might also permit an efficient classical description of the quantum evolution. This imprecise boundary is a consideration for quantum simulation algorithms  generally and remains a topic of active investigation.

Table \ref{table:algo_comparison} summarizes the asymptotic scaling of the number of circuits required per time step of both algorithms for open quantum system dynamical simulation on $n$ sites.

\begin{table}
\captionof{table}{Asymptotic number of circuits required per time step per Lindblad operator for both algorithms for an open system on $n$ sites. Here, $D$ is the domain size, and $I$ is is a subset of all $n$-bit strings for which the corresponding matrix elements are measured.\label{table:algo_comparison}
}
\begin{ruledtabular}
\begin{tabular}{ccc}
 Algorithm &\quad \# of qubits &\quad Circuits per Lindblad operator \\
 \hline
 I & $2n+1$ & $(n/D)4^D$ \\
 II & $n$ & $(n/D)4^D|I|^2$ \\
\end{tabular}
\end{ruledtabular}
\end{table}

\section{Results}
We demonstrate both algorithms on IBM Quantum hardware for two cases: the spontaneous emission of a two level system (TLS) in  a heat bath at zero temperature, and the dissipative transverse field Ising model (TFIM) on two sites. The TLS ($n=1$ from Table \ref{table:algo_comparison}) requires three physical qubits and one physical qubit to simulate with Algorithm I and II, respectively. The TFIM ($n=2$ from Table \ref{table:algo_comparison}) requires five and two physical qubits, respectively.

Considering Algorithm I, neither the TLS nor the two-site dissipative TFIM on 5 qubits have  constant depth circuit decompositions; Trotterizing both the real and imaginary time propagators results in a circuit with depth linear in the number of time steps. The resulting circuits are too deep for near-term devices. To overcome this limitation, we recompile the circuits as in Ref.~\cite{sun2021}. In all simulations, we correct for readout error using the built-in noise models in Qiskit \cite{Qiskit,temme2017error,kandala2019error,bravyi2020mitigating}. All measurements reported represent the average of 8192 shots and were repeated three times.   Sampling noise in the measurement of the expectation value of the Pauli strings can lead to numerical instabilities in the QITE linear system. Therefore,   when constructing the QITE matrix for Algorithm I, regularizers $1\times10^{-6}$ and 0.01, for the TLS and TFIM, respectively, were added to the diagonal terms   of the $S$ matrix   to increase the condition number of the matrix $S$ following the procedure in Ref.~\cite{motta2020}. No regularizers were used for Algorithm II.



\begin{figure}[ht]
  \centering
  {\includegraphics[width=\linewidth]{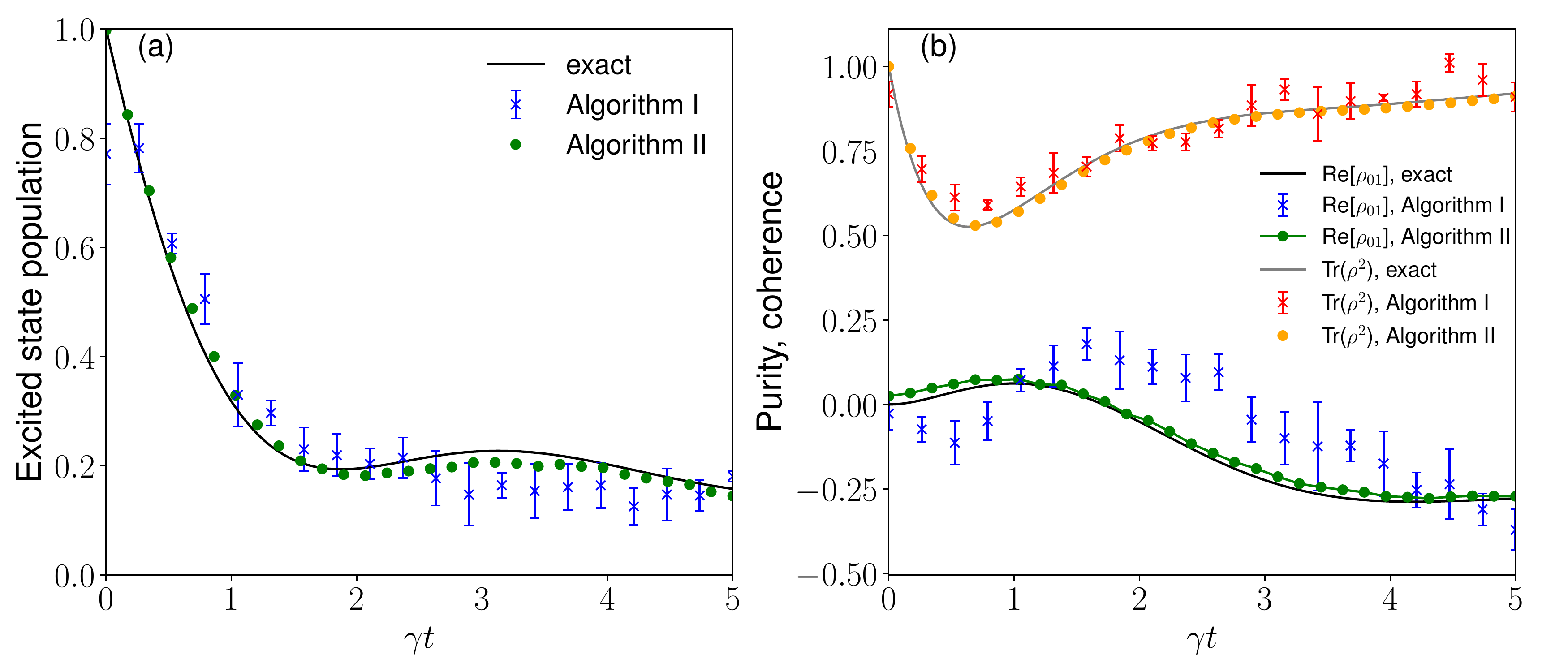}
   \phantomsubcaption\label{fig:2LS_population}
   \phantomsubcaption\label{fig:2LS_purity}}
  \caption{(a) Population of the excited state from numerical simulations obtained in QuTiP \cite{qutip1, qutip2} (black line), hardware using Algorithm I on $ibmq\_mumbai$  \cite{ibm} (blue crosses) and Algorithm II (green circles) on $ibmq\_casablanca$ \cite{ibm}. The deviation between the theoretical and experimental curves is largely due to gate error. The system approaches a  non-equilibrium steady state for $\gamma t\gtrsim5$. (b) Purity, $\tr{\rho^2}$ (grey line) and off-diagonal term, $\real{\rho_{10}}$ (black line), corresponding to non-diagonal observables obtained in in QuTiP \cite{qutip1, qutip2}. Hardware results are shown for Algorithm I (purity, red crosses; $\real{\rho_{10}}$, blue crosses) and for Algorithm II (purity, orange circles; $\real{\rho_{10}}$, green circles).
  Hardware results for the observable $\imag{\rho_{10}}$ agree with the exact solution similarly to $\real{\rho_{10}}$ but are omitted for clarity. For all hardware results for Algorithm I, the error bars are the standard deviation from three runs. The error bars for Algorithm II are smaller than the symbol size.}
  \label{fig:population-purity}
\end{figure}

We first present results for the TLS model with the Hamiltonian

\begin{equation}
    H=-\frac{\delta}{2}\sigma_z\ - \frac{\Omega}{2}\sigma_x
\end{equation} 
and the Lindblad operator $\sqrt{\gamma}\sigma_-$, where $\sigma_-$ is the lowering operator, $\delta$ is the detuning, $\Omega$ is the Rabi frequency, and $\gamma$ is the spontaneous emission rate. We consider here the overdamped case where $\gamma$ is on the order of the other energies in in the system. It was found via numerical simulations that to accurately capture the dynamics only the Pauli strings in the set $\left\{\sigma_x\otimes\sigma_z,\sigma_y\otimes\sigma_x,\sigma_y\otimes\sigma_z,\sigma_z\otimes\sigma_x\right\}$ needed to be included in the QITE unitary. 

We set $\delta=\Omega=\gamma=1$, and the initial state was chosen to be the excited state. In Fig.~\ref{fig:2LS_population}, we show the populations of the ground and excited states, with the experimental data averaged from three runs on IBM's $ibmq\_mumbai$ \cite{ibm} for Algorithm I and $ibmq\_casablanca$ \cite{ibm} for Algorithm II. Good qualitative agreement is obtained for all observables, with the deviation between the theoretical and experimental curves largely due to gate errors as confirmed by numerical simulations and noisy hardware emulations.

We observe an initial exponential decay in the population of the excited state due to spontaneous emission into the bath followed by an approach to the non-equilibrium steady steady state (NESS) for $\gamma t \gg 1$. Damped Rabi oscillations are visible between these two regimes. The populations in the NESS can be interpreted as a balance between the spontaneous emission due to coupling to the bath and the absorption and stimulated emission due to the Hamiltonian driving term $\sigma_x$ \cite{loudon2000quantum}. In the NESS, the combined spontaneous and stimulated emission rates are equal to the absorption rate. 

In the absence of driving by an external electric field ($\Omega=0$) the Hamiltonian is diagonal in the computational basis, resulting in the off-diagonal matrix elements $\rho_{01}=\cc{\rho_{10}}$ approaching zero as the system thermalizes. Figure ~\ref{fig:2LS_purity} shows that these matrix elements remain non-zero as the NESS is approached, indicating that the hardware correctly obtains the expected quantum coherence   as measured in the canonical basis.   Also shown in Fig.~\ref{fig:2LS_purity} is the purity $\tr{\rho^2}$, which does not correspond to a time-independent Hermitian observable on the system but can nonetheless be obtained from the density operator representation on the hardware. Time evolution preserves the inner product $\tr{\rho^2}=\braket{\rho}{\rho}$ on the quantum simulator, but the physical quantity, the normalized purity, $\tr{\rho^2}/\tr{\rho}$, is not constant.

  The larger deviation between the hardware results and the exact results for Algorithm I is attributed to the fact that Algorithm I is a three-qubit circuit requiring two-qubit gates, which are generally of lower fidelity than single qubit gates. Since Algorithm I for the TLS is a single qubit circuit, there are no infidelity contributions from two-qubit gates. In addition, the circuits required for Algorithm I are deeper than for Algorithm II, resulting in more gate errors. An additional breakdown of the error contributions due to hardware error and algorithmic error is provided in Fig. S1, in which  we compare the hardware results to noiseless numerical emulations.  

\begin{figure}[h]
  \centering
  \includegraphics[width=0.7\linewidth]{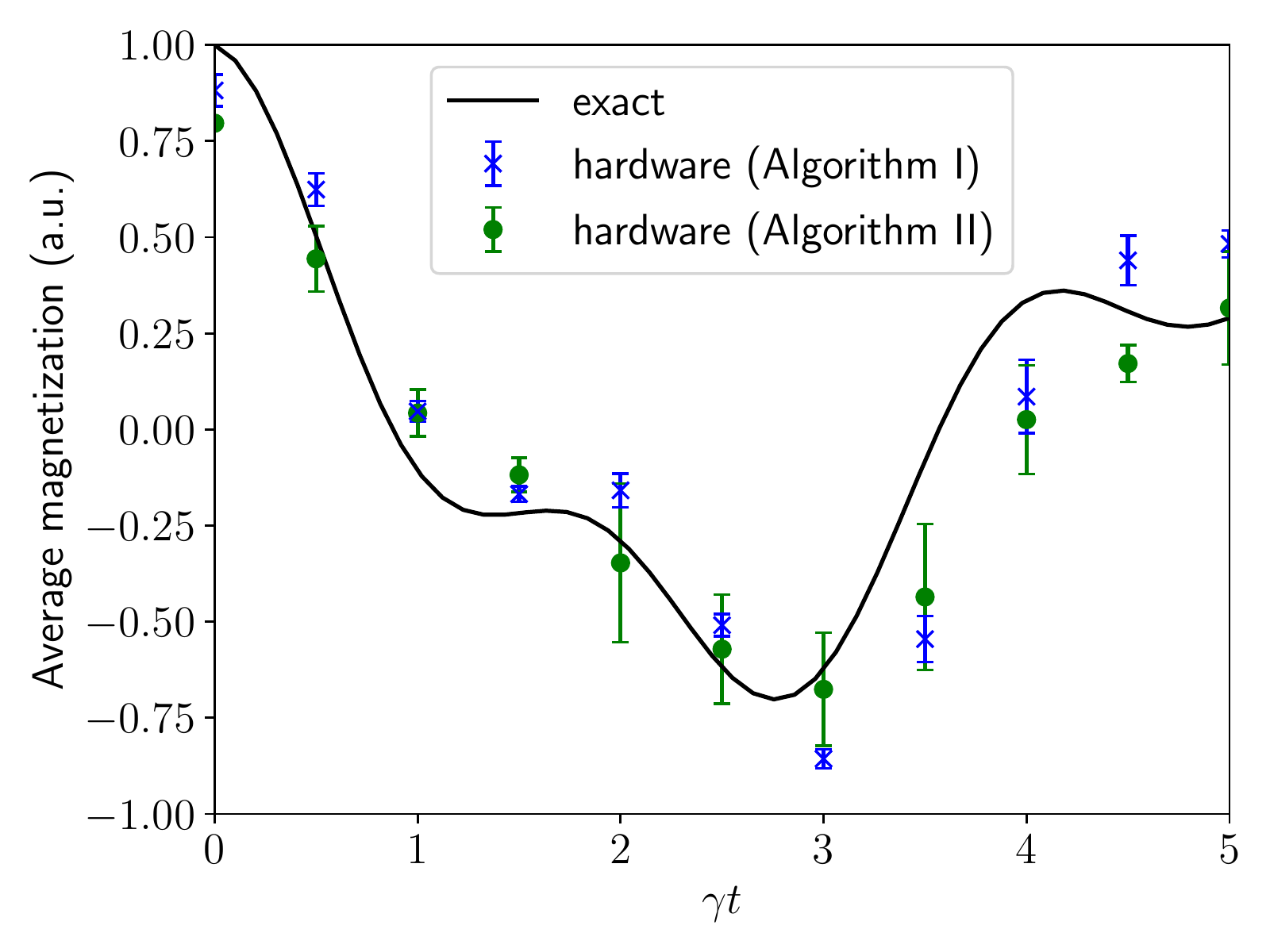}
  \caption{Average magnetization $N^{-1} \sum_i\langle Z_i\rangle$ for the dissipative transverse field Ising model on 2 sites (5 physical qubits for Algorithm I, 2 physical qubits for Algorithm II) using IBM Quantum's $ibmq\_guadalupe$ \cite{ibm} for Algorithm I (blue symbols), and $ibmq\_casablanca$ \cite{ibm} for Algorithm II (green symbols). Numerical solutions obtained in QuTiP are shown with black lines.  The error bars for both algorithms are the standard deviation from 3 hardware runs. Both algorithms qualitatively agree with the exact dynamics for all simulated times. The deviation between the hardware results and the exact result for Algorithm II is due mainly to Trotter gate error.}
  \label{fig:magnetization}
\end{figure}

We next present experimental and numerical results on the 2-site TFIM. The TFIM has the Hamiltonian 

\begin{equation}
    H=-J\sum_k\sigma_z^{(k)}\sigma_z^{(k+1)}-h\sum_k\sigma_x^{(k)}
\end{equation}  

\noindent and Lindblad operators $\sqrt{\gamma}\sigma_-^{(k)}$, with nearest neighbor coupling $J$, transverse magnetic field $h$, and decay rate $\gamma$. For this model, the number of required Pauli strings could not be reduced by symmetry in Algorithm I. To reduce circuit depth, 16 Pauli strings were randomly selected out of the 256 possible Pauli strings on 4 qubits to implement the QITE unitary. We chose 16 Pauli strings as a balance between too few Pauli strings, which results in a poor approximation to normalized imaginary time evolution, and too many Pauli strings, which results in a large computational overhead and an ill-conditioned QITE matrix. Increasing the number of Pauli strings does not qualitatively increase the accuracy, as shown in Fig. S2 of the Supplementary Materials.

Figure \ref{fig:magnetization} shows the average magnetization of the dissipative TFIM with the initial state given by both spins in the spin up state and $J=h=1$ and $\gamma=0.1$. Oscillations in magnetization are evident due to the relatively large transverse field. We observe qualitative agreement between the theoretical and experimental curves from Algorithm I with a Trotter step $\gamma t/N\sim 0.5$. For the small system size of 2 sites, all 4 bit-strings on 2 qubit were included in Algorithm II. Experimental results for Algorithm II are also in good qualitative agreement with the exact curve for all times. 

In Section \ref{sec:runtime}, we discussed the runtime and resources required by both algorithms in a general setting. We now discuss the relative computational cost required by each algorithm for the specific case of the 2-site TFIM hardware simulations. For the simulations considered here, Algorithm I is able to accurately describe the dissipative dynamics when using  16 out of the total of 256 Pauli strings.   Simulations of Algorithm I using up to 48 Pauli strings, shown in Fig. S2, show no significant increase in accuracy when using more than 16 Pauli strings.  In general, the number of required Pauli strings will be problem dependent.   Algorithm II requires measuring the matrix elements of all two-qubit Pauli strings at each time step, requiring 836 circuits per time step, versus only measuring expectation values of 16 operators in the case of Algorithm I, which requires 16 circuits. These measurements are only needed on a domain of fixed size. 

For larger dissipation rates $\gamma \sim J,h$, separate numerical simulations , presented in Fig S3,   show that both algorithms are able to accurately capture the magnetization dynamics. However, these simulations do not include the  error incurred by including a subset of bit-strings in Algorithm II. The actual algorithmic error of Algorithm II will thus depend on the accuracy of the representation of the density matrix with a subset of bit-strings for the given problem. Stochastic sampling of bit-strings may be a viable approach for larger systems. 

\section{Summary}
We have introduced digital quantum  algorithms for the time evolution of open quantum systems described by a Lindblad equation based on quantum imaginary time evolution. Algorithm I uses QITE to implement the non-unitary evolution introduced when the density operator is vectorized, whereas Algorithm II uses an adaptation of QITE to maintain a purification-based ansatz throughout the computation. Calculations for the spontaneous emission of a two level system and the dissipative transverse field Ising model, respectively, were carried out on IBM Quantum's quantum processors. Good   qualitative   agreement with the exact result was observed in all cases. These algorithms decrease the quantum resources required to simulate  open quantum systems governed by Lindblad master equations on quantum hardware.

\begin{acknowledgements}
H.~K., S.~N.~S. and A.~J.~M. were supported by NSF under Award No.~1839204.

\end{acknowledgements}

\clearpage

\bibliography{references}

\clearpage

\end{document}


\title{Supplemental Materials: Digital quantum simulation of open quantum systems using quantum imaginary time evolution}

\author{Hirsh Kamakari}
\affiliation{Division of Engineering and Applied Science, California Institute of Technology, Pasadena, CA 91125, USA}

\author{Shi-Ning Sun}
\affiliation{Division of Engineering and Applied Science, California Institute of Technology, Pasadena, CA 91125, USA}

\author{Mario Motta}
\affiliation{IBM Quantum, IBM Research Almaden, San Jose, CA 95120, USA}

\author{Austin J. Minnich}
\email{aminnich@caltech.edu}
\affiliation{Division of Engineering and Applied Science, California Institute of Technology, Pasadena, CA 91125, USA}

\date{\today}

\maketitle

\setcounter{equation}{0}
\setcounter{figure}{0}
\setcounter{table}{0}
\setcounter{page}{1}
\makeatletter
\renewcommand{\theequation}{S\arabic{equation}}
\renewcommand{\thefigure}{S\arabic{figure}}
\renewcommand{\bibnumfmt}[1]{[S#1]}
\renewcommand{\citenumfont}[1]{S#1}


\section{Derivation of the QITE linear system for Algorithm II}
\label{ap:cj_derivation}
Here we derive the QITE linear systems which need to be solved to obtain the time evolution of the density operator ansatz. Consider $\ket{\rho} = \sum_xp_xT\ket{x}\otimes \cc{T}\ket{x}$, with $T$ unitary. The complex time propagator is the same as in the vectorization method, 
\begin{equation}
X(t) = \exp{\left(\left[-i\identity\otimes H + iH^\top\otimes\identity + \sum_{k}(\cc{L_k}\otimes L_k -\frac{1}{2}\identity\otimes(L_k^\dagger L_k) -\frac{1}{2}(L_k^\top \cc{L_k})\otimes\identity)\right]t\right)}.
\end{equation}
Trotterizing results in
\begin{equation}
{
\begin{split}
&X(t) =\\
&\left\{\left[\exp{\left(iH^\top\tau\right)}\otimes\exp{\left(-iH\tau\right)}\right]\prod_k\left[\exp{\left(- \frac{L_k^\top \cc{L_k}\tau}{2} \right)}\otimes\exp{\left(- \frac{L_k^\dagger L_k\tau}{2} \right)}\right]\exp{(\cc{L_k}\otimes L_k\tau)}\right\}^n \\
&+ \bigo{\tau^2n}
\end{split}
}
\end{equation}
with $\tau = t/n$. Using the identity $\cc{\exp{(-iA)}}=\exp{(iA^\top)}$ for $A$ Hermitian and $\cc{\exp{(B)}}=\exp{(\cc{B})}$ for arbitrary $B$, the propagator can be rewritten as 
\begin{equation}
X(t) = \left\{[U\otimes \cc{U}]\prod_k[V_k\otimes \cc{V_k}]W_k\right\}^n 
+ \bigo{\tau^2n}
\end{equation}
with $U:=\exp{(iH^\top\tau)}$, $V_k:=\exp{(-L_k^\top \cc{L_k}\tau/2)},$ and $W_k:=\exp{(\cc{L_k}\otimes L_k\tau)}$. It is immediate that evolution with $U\otimes\cc{U}$ preserves the ansatz, as $(U\otimes\cc{U})\sum_xp_xT\ket{x}\otimes \cc{T}\ket{x}=\sum_xp_x(UT)\ket{x}\otimes (\cc{UT})\ket{x}$. The term $V\otimes\cc{V}$ also preserves the ansatz, but is an imaginary time evolution with Hamiltonian $L_k^\top \cc{L_k}/2$, and so requires a modified QITE algorithm, described below, for implementing $\exp{(-L_k^\top \cc{L_k}\tau/2)}T\ket{x}$. Due to the non-unitarity of $V_k$, we expect that in addition to a unitary evolution of the state, the weights $p_x$ will also evolve in time. The final term, $W_k=\exp{(\cc{L_k}\otimes L_k\tau)}$, does not preserve the ansatz, and we use a modified version of QITE, described below, to effectively apply $W_k$ while preserving the form of ansatz.
 
\subsection{Implementing $V_k\otimes\cc{V_k}$ via a QITE adaptation}
Under real time evolution by the non-unitary operator $V_k\otimes\cc{V_k}$, the evolution of $\ket{\rho}$ can be expressed as
\begin{equation}
    V_k\otimes\cc{V_k}\sum_xp_xT\ket{x}\otimes \cc{T}\ket{x}=\sum_x(p_x+q_x)\exp{(iA)}T\ket{x}\otimes \exp{(-i\cc{A})}\cc{T}\ket{x} + O(\tau^2),
\end{equation}
with $q_x\in\reals$ and $A$ a Hermitian operator with $\dnorm{A}_2=\bigo{\tau}$. Defining $B:=(1/2)\transpose{L_k}\cc{L_k}$, we then have, to first order in $\tau$,
\begin{equation}
    \exp{(-\tau B)}\otimes \exp{(-\tau \cc{B})}\sum_xp_xT\ket{x}\otimes\cc{T}\ket{x}=\sum_x(p_x+q_x)\exp{(iA)}T\ket{x}\otimes \exp{(-i\cc{A})}\cc{T}\ket{x}.
\end{equation}
Expanding both sides to first order in $\tau$ and discarding higher order terms results in 
\begin{equation}
    \begin{split}
    -\tau\sum_xp_x(BT\ket{x}\otimes \cc{T}\ket{x}+T\ket{x}\otimes \cc{B}\cc{T}\ket{x})&=\\ \sum_xq_xT\ket{x}\otimes \cc{T}\ket{x}+i\sum_x&p_x(AT\ket{x}\otimes \cc{T}\ket{x}-T\ket{x}\otimes \cc{A}\cc{T}\ket{x}).
    \end{split}
    \label{eqn:type1q}
\end{equation}
Taking the inner product of Eq. (\ref{eqn:type1q}) with $\bra{y}T^\dagger\otimes\bra{y}\transpose{T}$ results in 
\begin{equation}
    \begin{split}
        -\tau\sum_xp_x(\bra{y}T^\dagger BT\ket{x}\bra{y}\transpose{T} \cc{T}\ket{x}+\bra{y}T^\dagger T\ket{x} \bra{y}\transpose{T}\cc{B}\cc{T}\ket{x})&=\\ \sum_xq_x\bra{y}T^\dagger T\ket{x}\bra{y}\transpose{T}\cc{T}\ket{x}+i\sum_xp_x(\bra{y}T^\dagger AT\ket{x}\bra{y}\transpose{T}\cc{T}\ket{x}&-\bra{y}T^\dagger T\ket{x}\bra{y}\transpose{T}\cc{A}\cc{T}\ket{x}).
    \end{split}
\end{equation}
Using the identities $\bra{y}T^\dagger T\ket{x}=\bra{y}\transpose{T}\cc{T}\ket{x}=\delta_{xy}$ for $T$ unitary results in
\begin{equation}
    -\tau p_y(\bra{y}T^\dagger BT\ket{y}+ \bra{y}\transpose{T}\cc{B}\cc{T}\ket{y})= q_y+ip_y(\bra{y}T^\dagger AT\ket{y}-\bra{y}\transpose{T}\cc{A}\cc{T}\ket{y}).
    \label{eqn:q_deriv}
\end{equation}
Because $A$ is Hermitian, we additionally have $\bra{y}T^\dagger A T\ket{y} = \bra{y}\transpose{T}\cc{A}\cc{T}\ket{y}$ so that the last two terms on the right-hand side cancel, resulting in 
\begin{equation}
        q_y=-2\tau p_y\real{\bra{y}T^\dagger BT\ket{y}}
\end{equation}

With the $q_x$'s determined, we can now determine the operator $A$. Rearranging Eq. (\ref{eqn:type1q}), we first isolate the terms containing $A$:
\begin{equation}
    \begin{split}
    i\sum_xp_x(AT\ket{x}\otimes \cc{T}\ket{x}-T\ket{x}\otimes \cc{A}\cc{U}\ket{x}) &= \\
    -\tau\sum_xp_x(BT\ket{x}\otimes \cc{T}\ket{x}&+T\ket{x}\otimes \cc{B}\cc{T}\ket{x}) -  \sum_xq_xT\ket{x}\otimes \cc{T}\ket{x}.
    \end{split}
    \label{eqn:type1A}
\end{equation}
We define the right hand side as 
\begin{equation}
    \ket{\Phi}=-\tau\sum_xp_x(BT\ket{x}\otimes \cc{T}\ket{x}+T\ket{x}\otimes \cc{B}\cc{T}\ket{x}) -  \sum_xq_xT\ket{x}\otimes \cc{T}\ket{x}.
\end{equation}
We then decompose $A$ into a sum over Pauli strings with domain size $D$, $A=\sum_ja_j\sigma_j$, where the $\sigma_j$ are Pauli strings acting on  at most $D$ qubits, $a_j\in\reals$ and $a_j=\bigo{\tau}$ for all $j$. Substituting into the left hand side of Eq. (\ref{eqn:type1A}) yields 
\begin{equation}
    i\sum_{x,j}p_xa_j(\sigma_jT\ket{x}\otimes \cc{T}\ket{x}-T\ket{x}\otimes\cc{\sigma_j}\cc{T}\ket{x})=\sum_ja_j\ket{v_j}=\ket{\Phi},
\end{equation}
where we have defined the vectors $\ket{v_j}:=\sum_xp_x(\sigma_jT\ket{x}\otimes T^*\ket{x}-T\ket{x}\otimes\sigma_j^*T^*\ket{x})$. Denoting by $f$ the function \begin{equation}
    f(a)=\dnorm{\ket{\Phi}-i\sum_ja_j\ket{v_j}}^2=\braket{\Phi}{\Phi}+i\sum_j(a_j^*\braket{v_j}{\Phi}-a_j\braket{\Phi}{v_j})+\sum_{jk}a_j^*a_k\braket{v_j}{v_k}),
\end{equation} the optimal coefficients $a_j$ are determined by minimizing $f$. This results in the set of equations
\begin{equation}
    0=\frac{\partial f}{\partial a_k}=-\imag{\braket{v_k}{\Phi}}+\sum_ja_j\real{\braket{v_k}{v_j}}.
\end{equation}
Defining the matrix $S_{jk}:=\real{\braket{v_j}{v_k}}$ and the vector $b_j:=\imag{\braket{v_j}{\Phi}}$, the optimal coefficients $a$ are the solution to the linear system $Sa=b$.

Using the definition's of $\ket{v_j}$ and $\ket{\Phi}$, we calculate calculate the matrix elements of $S$ as
\begin{equation}
    S_{jk}=\real{\braket{v_j}{v_k}}=\sum_xp_x^2\real{\bra{x}T^\dagger(\sigma_j\sigma_k+\sigma_k\sigma_j)T\ket{x}}-2\sum_{xy}p_xp_y\real{\bra{x}T^\dagger\sigma_jT\ket{y}\bra{x}T^\dagger\sigma_kT\ket{y}},
\end{equation}
and the elements of $b$ as
\begin{equation}
    b_j=-2\tau\left(\sum_xp_x^2\imag{\bra{x}T^\dagger \sigma_jBT\ket{x}}+\sum_{xy}p_xp_y\imag{\bra{x}T^\dagger\sigma_jT\ket{y}\bra{y}T^\dagger B^\dagger T\ket{x}}\right).
\end{equation}

\subsection{Implementing $W_k$ via a QITE adaptation}
The real time evolution corresponding to $W_k=\exp{(\tau\cc{L_k}\otimes L_k)}$ can be determined completely analogously to that of $V_k\otimes\cc{V_k}$ above. The resulting equations are 
\begin{equation}
\begin{cases}
    q_y=\tau\sum_xp_x\snorm{\bra{y}T^\dagger L_kT\ket{x}}^2\\
    S_{jk}=\real{\braket{v_j}{v_k}}\\
    b_j=\imag{\braket{v_j}{\Phi}}
\end{cases}
\end{equation} 
where $Sa=b$ gives the optimal Pauli strings. The matrix elements for $S$ are the same, as the vectors $\ket{v_j}$ are identical in both cases. Since $\ket{\Phi}$ has a different form, the elements of $b$ are modified and given by
\begin{equation}
    b_j=\imag{\braket{v_j}{\Phi}}=2\tau\sum_{xy}p_xp_y\imag{\bra{x}T^\dagger\sigma_jL_kT\ket{y}\bra{y}T^\dagger L_k^\dagger T\ket{x}}.
\end{equation}

\subsection{Conservation of Probability}
The trace of the density operator, given by $\tr{\rho}=\sum_xp_x=1$, is preserved by time evolution generated by the Lindblad equation. Here we show that time evolution via Algorithm II also maintains the trace. The trace is preserved if the sum of all $q_x$'s is zero at each time step. This requires summing the contributions to the $q_x$'s from both the $V_k$ and $W_k$ terms as follows:
\begin{align}
    \label{eqn:sum_q_0}
    \begin{split}
    \sum_yq_y &= \sum_y\left(-\tau p_y\real{\bra{y}T^\dagger L_k^\dagger L_kT\ket{y}}+\tau\sum_xp_x\snorm{\bra{y}T^\dagger L_kT\ket{x}}^2\right)\\
    &= -\tau\sum_yp_y\bra{y}T^\dagger L_k^\dagger L_kT\ket{y}+\tau\sum_xp_x\sum_y\bra{x}T^\dagger L_k^\dagger T\ket{y}\bra{y}T^\dagger L_k T\ket{x}\\
    &= -\tau\tr{\rho L_k^\dagger L_k}+\tau\sum_xp_x\bra{x}T^\dagger L_k^\dagger L_kT\ket{x}\\
    &= -\tau\tr{\rho L_k^\dagger L_k} + \tau\tr{\rho L_k^\dagger L_k}\\
    &= 0
    \end{split}
\end{align}

\subsection{Measuring Observables} 
The result of the above real and imaginary time evolution, the weights $p_x(t)$ and the Hermitian operator $A(t)$ can be used to calculate the expectation value of any observable. Inverting the Choi-Jamio\l kowski isomorphism gives us the density operator $\rho(t)=\sum_xp_x(t)T(t)\ket{x}\bra{x}T^\dagger(t)$. Observables $O$ are then calculated as
\begin{equation}
\begin{split}
    \qexpect{O(t)}=\tr{O\rho(t)}&=\sum_{xy}p_x(t)\bra{y}OT(t)\ket{x}\bra{x}T^\dagger(t)\ket{y}\\
    &=\sum_{xy}p_x\bra{x}T^\dagger\ket{y}\bra{y}OT\ket{x}\\
    &=\sum_{x}p_x\bra{x}T^\dagger\left(\sum_y\ket{y}\bra{y}\right)OT\ket{x}\\
    &=\sum_xp_x(t)\bra{x}T^\dagger(t)OT(t)\ket{x}.
\end{split} 
\end{equation}
Beyond a certain number of qubits, storing all the $p_x(t)$'s is not possible, and a stochastic sampling approach is needed. Locality conditions suggest one possible approach to efficient sampling, described at the end of section (\ref{subsec:matrix_elements}), which converges faster than uniform random sampling.

\subsection{Measuring Matrix Elements}
\label{subsec:matrix_elements}
To obtain the coefficients $q_x$ and $a_i$, we need to measure various matrix elements. In general, we can decompose any operator into a sum over Pauli strings, $X=\sum_jx_i\sigma_i$. Since $\bra{x}X\ket{y}=\sum_jx_i\bra{x}\sigma_i\ket{y}$, we then need to measure $\bra{x}\sigma_i\ket{y}$ for all Pauli strings $\sigma_i$. This can be done using the following identities:
\begin{align}
    2\real{\bra{x}X\ket{y}}&=\dfrac{\bra{x}+\bra{y}}{\sqrt{2}}X\dfrac{\ket{x}+\ket{y}}{\sqrt{2}}-\dfrac{\bra{x}-\bra{y}}{\sqrt{2}}X\dfrac{\ket{x}-\ket{y}}{\sqrt{2}},\\
    2\imag{\bra{x}X\ket{y}}&=\dfrac{\bra{x}+i\bra{y}}{\sqrt{2}}X\dfrac{\ket{x}-i\ket{y}}{\sqrt{2}}-\dfrac{\bra{x}-i\bra{y}}{\sqrt{2}}X\dfrac{\ket{x}+i\ket{y}}{\sqrt{2}}.
\end{align}

In general, the state $(\ket{x}+ i^p  \ket{y})/\sqrt{2}$, with $p \in \{0,1,2,3\}$, requires a quantum circuit 
comprising $m$ CNOT gates and having depth $m+1$, where $m$ is the
Hamming distance between the binary strings $x,y$ \cite{parrish2019quantum}.
Indeed, one can find an index $k$ such that $x_k \neq y_k$. Without
loss of generality, one can assume that $x_k = 1$ (otherwise, just
invert the roles of $x$, $y$ and replace $p$ with $-p$  mod $4$).
One can then define the sets $S = \{ l : x_l = 1 , l \neq k \}$, 
$T = \{ l : x_l \neq y_l , l \neq k \}$. 
Finally, starting from a register of $n$ qubits prepared in 
$|0\rangle^{\otimes n}$, the desired state is obtained by: 
(i) applying a product of $X$ gates on qubits in the set $S$, 
$\prod_{l \in S} X_l$,
(ii) applying to qubit $k$ the gate $g_p = H, SH, ZH, ZSH$, for $p=0,1,2,3$, respectively, and
(iii) applying a product of CNOT gates to qubits in $T$ controlled
by qubit $k$, $\prod_{l \in S} c_k X_l$.

For local observables the state preparation is simpler, as described in the following. Consider a $k$-qubit observable $X^{(k)}\otimes\identity_{n-k}$, with $X^{(k)}$ acting non-trivially on $k$ qubits out of a total of $n$ qubits. Then
\begin{align}
    \bra{x}X^{(k)}\otimes\identity_{n-k}\ket{y}&=\bra{x_1,\dots,x_k,x_{k+1},\dots,x_n}X^{(k)}\otimes\identity_{n-k}\ket{y_1,\dots,y_k,y_{k+1},\dots,y_n}\\
    &=\delta_{x_{k+1},y_{k+1}}\cdots\delta_{x_n,y_n}\bra{x_1,\dots,x_k}X^{(k)}\ket{y_1,\dots,y_k}.
\end{align}
Thus we need only to prepare the states 
\begin{equation}
\begin{split}
    \frac{\ket{x_1,\dots,x_k,x_{k+1},\dots,x_n}+\ket{y_1,\dots,y_k,x_{k+1},\dots,x_n}}{\sqrt{2}}&=\\
    \frac{\ket{x_1,\dots,x_k}+\ket{y_1,\dots,y_k}}{\sqrt{2}}&\otimes\ket{x_{k+1},\dots,x_n}.
\end{split}
\label{eq:local_matrix_elements}
\end{equation}
Since $k$ is typically small, only 1 or 2 qubits in most cases and independent of the system size, this state can be efficiently prepared. The form of Eq.~(\ref{eq:local_matrix_elements}) suggests a stochastic sampling method to determine which $p_x$'s to store classically. For simplicity we describe the case of qubits in a line, and the indices $1,\dots,n$ labelling the sites with the observable acting on the first $k$ qubits. The general case is similar. Since the matrix elements will depend more heavily on qubits $1,\dots,k+m$ for some cutoff $m$, we can sample with higher frequency on the first $k+m$ qubits and with lower frequency on the rest.  In addition, in many cases we expect the dissipation channels $L_k$ to reduce long range correlations, further increasing the convergence rate of local sampling.

\newpage
\section{Hardware errors in the two level system simulation}
\begin{figure}[!htb]
  \centering
  \includegraphics[width=0.6\linewidth]{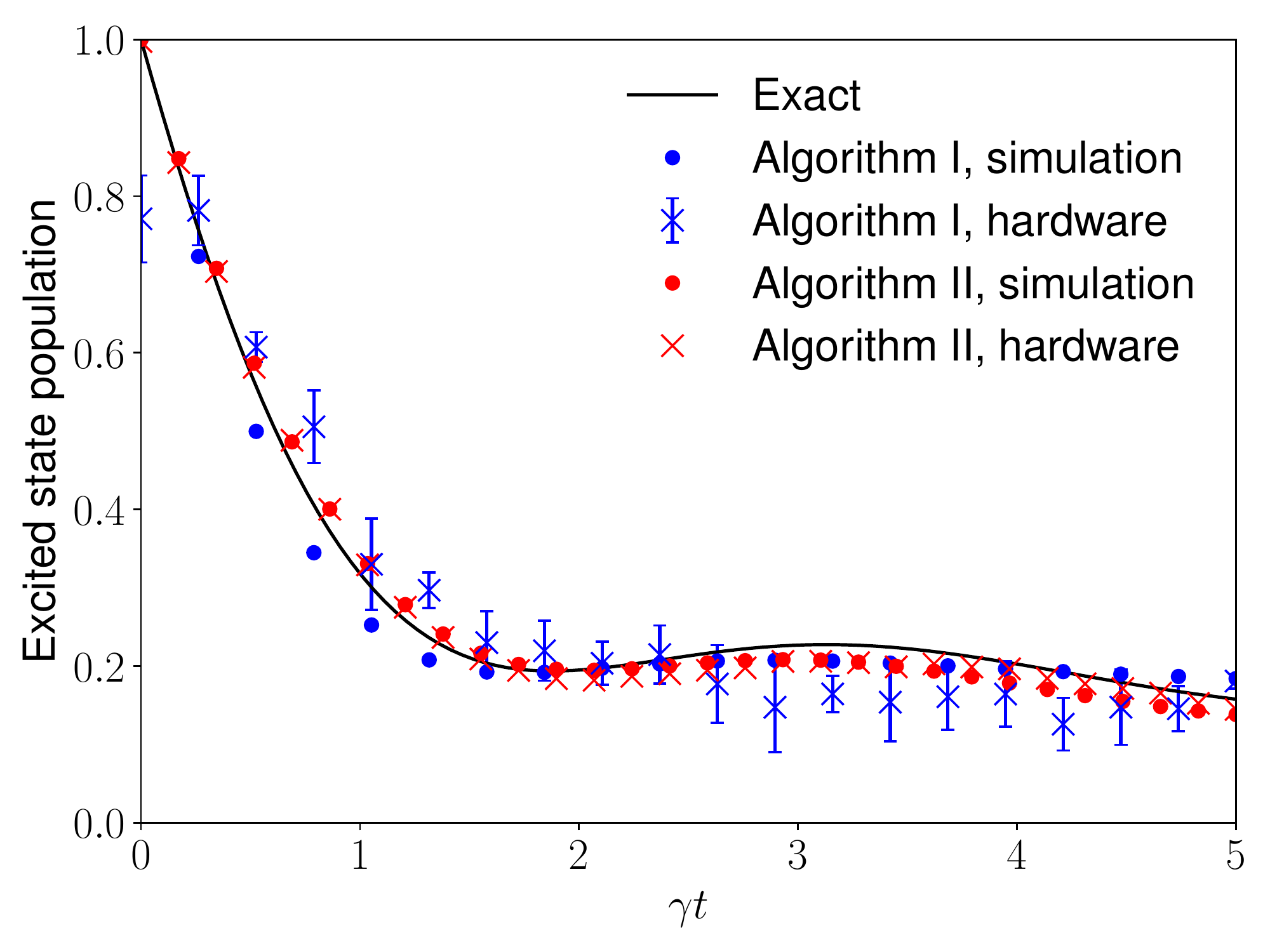}
  \caption{Excited state population for the two level system (TLS). The solid curve is the exact solution and the blue and red dots are noiseless numerical emulations of Algorithm I and II, respectively. The blue and red crosses are the hardware results presented in the main text for Algorithm I and II, respectively. The deviation between hardware and simulation results for Algorithm I are larger than for Algorithm II, which we attribute to hardware error resulting from the larger circuit depth and number of qubits needed for the Algorithm I.}
  \label{fig:trotter_test}
\end{figure}

We compare the hardware results to noiseless numerical emulations in Fig. \ref{fig:trotter_test} so as to further separate hardware and algorithmic errors. The noiseless numerical emulations were run with the same circuits as in the hardware trials but using IBM's \textit{qasm\_simulator}. From Fig. \ref{fig:trotter_test}, we see that Algorithm I has a larger deviation between the emulation and hardware data. This difference can be accounted for by the fact that the hardware experiment for Algorithm II requires only a single qubit, so the density matrix for all time steps can be obtained from only single qubit rotations. Single qubit simulations can always be compiled to a constant depth regardless of the number of time steps, resulting in lower depth circuits and correspondingly lower total gate error. In addition, since 2-qubit gates are generally lower fidelity than single-qubit gates, there is no infidelity contribution due to 2-qubit gates in Algorithm II.

\newpage
\section{Number of Pauli strings in the transverse field Ising model}

\begin{figure}[!htb]
  \centering
  \includegraphics[width=0.7\linewidth]{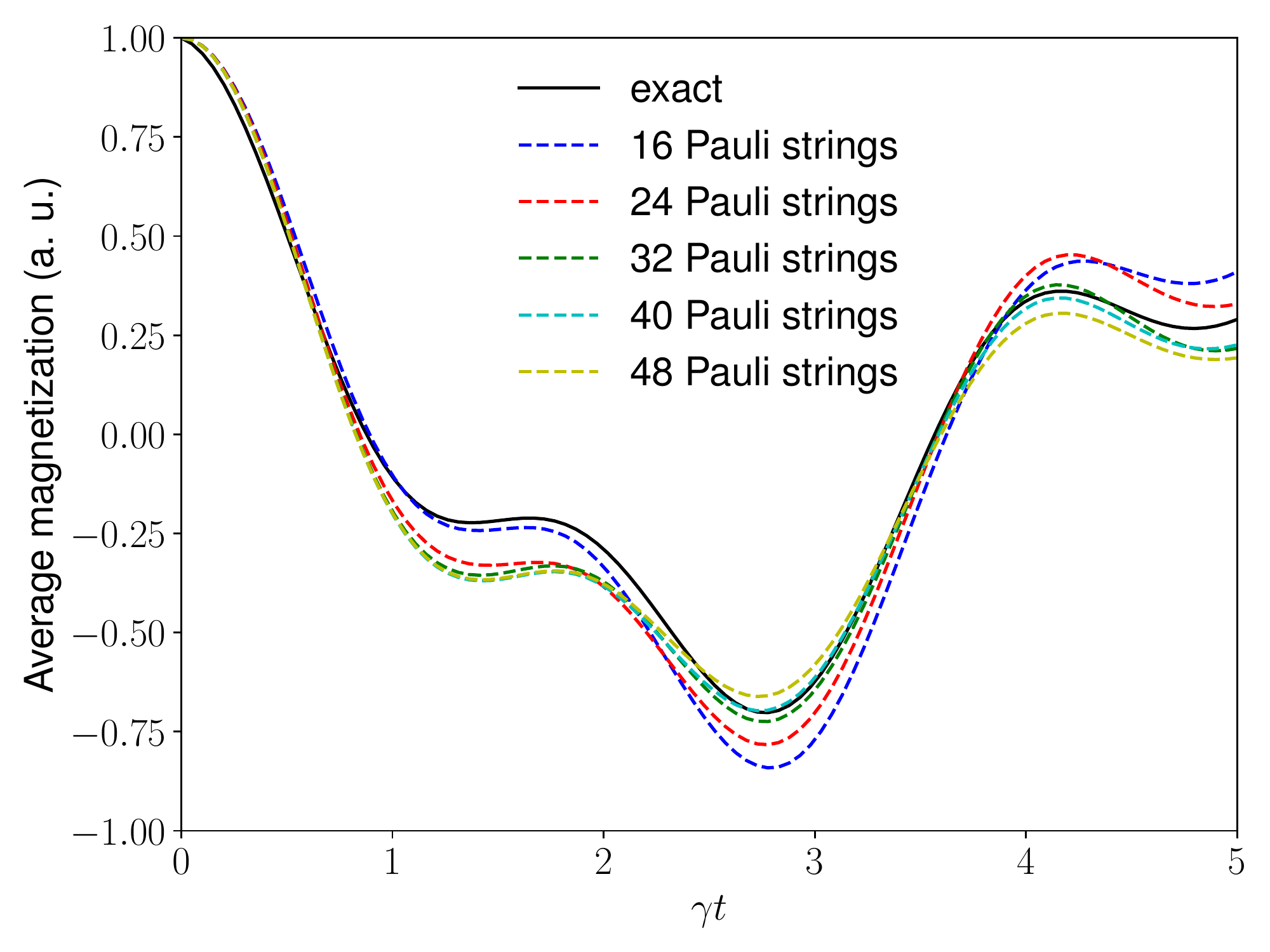}
  \caption{Noiseless numerical simulations for the transverse field Ising model (TFIM) using Algorithm I with increasing number of Pauli strings included. Here the dissipation rate is $\gamma=0.1$. The black solid curve is the exact result, and the blue dashed curve is a simulation of Algorithm I using the same 16 Pauli's as in the main text. The red, green, light blue, and yellow dashed curves are noiseless numerical simulations of Algorithm I obtained from including an increasing number of Pauli strings in the simulation. From these simulations we see that only marginal increase in accuracy is obtained from including a larger number of Pauli strings.}
  \label{fig:pauli_string_test}
\end{figure}

Exactly simulating the 2-site TFIM using Algorithm I requires measuring expectation values of the 256 Pauli strings on 4 qubits. To reduce the runtime of the Algorithm, we use a subset of all Pauli strings. We show in Fig.~\ref{fig:pauli_string_test} that increasing the number of included Pauli strings beyond 16 has only a minor effect on the observables.
 
\newpage
\section{Effect of dissipation rate on algorithm precision}
\begin{figure}[h]
  \centering
  \includegraphics[width=\linewidth]{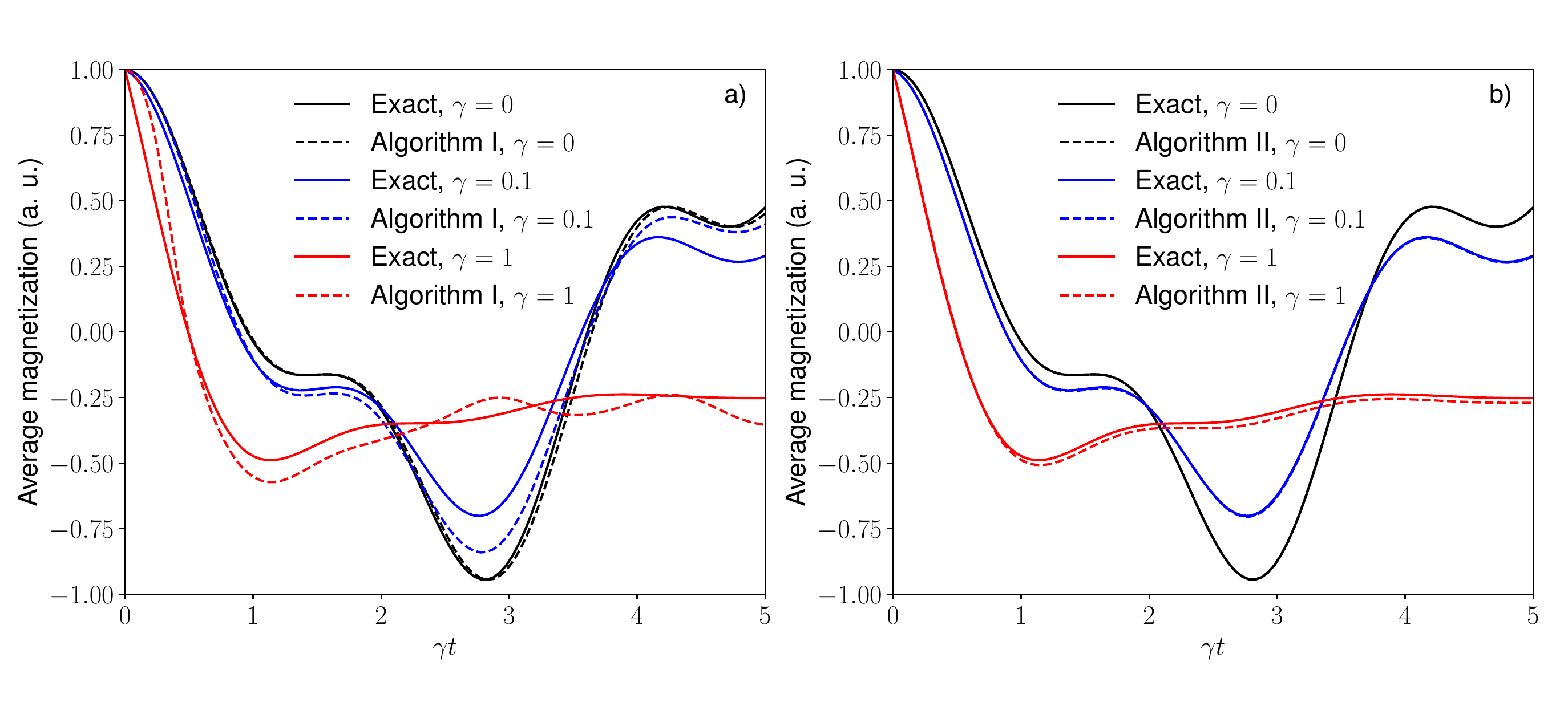}
  \caption{The effect of increasing the dissipation rate from $\gamma=0$ to $\gamma=1$. a) Noiseless simulation of Algorithm I using 16 Pauli strings. The same qualitative error is obtained for all dissipation rates simulated. b) Noiseless simulation of Algorithm II.}
  \label{fig:gamma_test}
\end{figure}

Here we study the effects of increasing dissipation rates on the accuracy of both algorithms. In general, we should expect larger algorithmic errors when larger dissipation rates are simulated since both the Trotter error and QITE error increase with the operator norm of the Lindblad operators. Larger dissipation rates correspond to Lindblad operators with larger norms and hence larger algorithmic errors. To understand how increasing dissipation rates affect both algorithm's errors, we performed simulations of the 2-site TFIM with dissipation rates ranging from $\gamma=0$ to $\gamma=1$. Figure  \ref{fig:gamma_test} shows the results of the simulations. We see that in this specific case, which includes 16 Pauli strings for Algorithm I and all possible bit-strings for Algorithm II, Algorithm II performs qualitatively better than Algorithm I for all dissipation rates. Although Algorithm II performs better for the simulations shown in Fig.~\ref{fig:gamma_test}, we have not considered the error due to bit-string selection. For larger systems where all bit-strings cannot be included, there will be additional errors introduced by including only a strict-subset of all possible bit-strings.

